\documentclass[twocolumn,aps,prl,amsmath,amssymb,superscriptaddress,bibnotes]{revtex4}
\usepackage{amsmath}
\usepackage{rotating}
\usepackage{extarrows}
\usepackage[
            colorlinks=true,
            linkcolor=blue,
            citecolor=blue,
            urlcolor=blue
            ]{hyperref}
\usepackage{graphicx}
\usepackage{dcolumn}
\usepackage{bm}
\usepackage[all]{xy}
\usepackage{indentfirst}
\usepackage{amsmath}
\usepackage{bbm}
\usepackage{color}
\usepackage{multirow}
\usepackage{mathrsfs}
\usepackage{euscript}
\usepackage{amssymb}
\usepackage{floatrow}
  \newfloatcommand{capbtabbox}{table}[][\FBwidth]
\newcommand{\expe}[1]{\langle #1 \rangle}

\newcommand{\ket}[1]{|#1\rangle}
\newcommand{\bra}[1]{\langle #1|}

\begin{document}
\title{Simultaneous Verification of Genuine Multipartite Nonlocality and Full Network Nonlocality}

\author{Ning-Ning Wang$^{1,2,3,5\,*}$, 
Xue Yang$^{4}$\footnote{These authors contribute equally.},
Yan-Han Yang$^{4}$,
Chao Zhang$^{1,2,3,5}$\footnote{drzhang.chao@ustc.edu.cn}, 
Ming-Xing Luo$^{4,5}$\footnote{mxluo@swjtu.edu.cn}, 
Bi-Heng Liu$^{1,2,3,5}$, 
Yun-Feng Huang$^{1,2,3,5}$\footnote{hyf@ustc.edu.cn}, 
Chuan-Feng Li$^{1,2,3,5}$\footnote{cfli@ustc.edu.cn}, 
Guang-Can Guo$^{1,2,3,5}$}

\affiliation{1. CAS Key Laboratory of Quantum Information, University of Science and Technology of China, Hefei, 230026, P. R. China
\\
2. Anhui Province Key Laboratory of Quantum Network, University of Science and Technology of China, Hefei 230026, P. R. China
\\
3. CAS Center For Excellence in Quantum Information and Quantum Physics, University of Science and Technology of China, Hefei, 230026, P. R. China
\\
4. School of Information Science and Technology, Southwest Jiaotong University, Chengdu 610031, P. R. China
\\
5. Hefei National Laboratory, University of Science and Technology of China, Hefei, 230088, P. R. China }

\begin{abstract}

Genuine multipartite nonlocality and nonlocality arising in networks composed of several independent sources have been separately investigated. While some genuinely entangled states cannot be verified by violating a single Bell-type inequality, a quantum network consisting of different sources allows for the certification of the non-classicality of all sources. In this paper, we propose the first method to verify both types of nonlocality simultaneously in a single experiment. We consider a quantum network comprising a bipartite source and a tripartite source. We demonstrate that there are quantum correlations cannot be simulated if the tripartite source distributes biseparable systems while the bipartite source distributes even stronger-than-quantum systems. These correlations can be used to verify both the genuine multipartite nonlocality of generalized Greenberger-Horne-Zeilinger states and the full network nonlocality that is stronger than all the existing results. Experimentally, we observe both types of nonlocality in a high fidelity photonic quantum network by violating a single network Bell inequality. 

\end{abstract}

\maketitle

{\it Introduction.--}Bell's theorem has revolutionized our understanding of entangled quantum systems by revealing that they can exhibit correlations that defy the constraints of any local realistic theory \cite{Bell}. This groundbreaking discovery offers a new perspective on the challenge initially posed by the Einstein-Podolski-Rosen (EPR) argument \cite{EPR}.  Bell nonlocality, which transcends EPR's explanation of steering, sheds light on the profound quantum effects inherent in general entangled states \cite{Gisin1991,Gisin1992}. This phenomenon has been empirically confirmed in numerous Bell experiments \cite{PhysRevLett.115.250401,PhysRevLett.115.250402,Hensen2015}, where violations of the Clauser-Horne-Shimony-Holt (CHSH) inequality \cite{CHSH} have been reported. Beyond its fundamental significance, Bell nonlocality has also proven to be a crucial resource in device-independent quantum information tasks \cite{Acin2007,Pironio2010}.

The first extension of bipartite nonlocality is to characterize the global correlations of multipartite entanglement. The genuine multipartite nonlocality (GMN) specifically pertains to the correlations observed in systems with multiple particles, where these correlations cannot be simulated using classical mixtures of bipartite separable correlations under the no-signaling principle \cite{Svetlichny,Popescu1992}. This distinction highlights the fundamentally distinct and complex nature of nonlocality in multipartite entangled systems. The verification of the GMN hinges on the utilization of specialized Bell-type inequalities, such as the Mermin-Ardehali-Belinskii-Klyshko (MABK) and Svetlichny inequalities \cite{Mermin,Ardehali,Belinskii,Svetlichny}. Other methods have been proposed to examine specific entangled states within this context \cite{Chen,Yu}. Notably, Scarani and Gisin \cite{SG} have contributed to this area by demonstrating that generalized Greenberger-Horne-Zeilinger (GHZ) states \cite{GHZ} do not violate the MABK inequalities \cite{Mermin,Ardehali,Belinskii}. This intriguing result has been further corroborated for general states \cite{Zukowski2002,CL,RD}. These findings have sparked renewed interest in the exploration of novel methods to verify GMN \cite{Contreras2021,Contreras2022}, presenting an intellectually stimulating and compelling problem for further investigations within the realm of quantum nonlocality \cite{Gallego2012,Xavier2021}.

Another intriguing extension involves relaxing the assumption of sources, leading to the exploration of nonlocality in network scenarios. Specifically, various results have focused on investigating nonlocal correlations arising from entanglement-swapping experiments in networks \cite{BGP,Rosset2016,Chaves2016,Luo2018,Renou2019,Alejandro2019,Ivan2020,Patricia2021,Tavaki2021,Renou2022,Alejandro2023,Gatto2023,TPLR}, where independent parties can become entangled despite a lack of causal connections. Beyond the single local hidden variable (LHV) assumption in standard Bell tests, it is more natural to consider models with independent hidden variables that replicate the network structure. This has led to the study of bilocal hidden variable models and the associated phenomenon of non-bilocality. However, this kind of nonlocality may be simulated by mixing the classical correlations and nonlocal correlations or post-quantum correlations \cite{Alej2022}. A further extension is then to rule out all of these correlations by defining the stronger local model, that is, any hybrid network consisting of at least one classical source, while all the others can be any no-signaling sources. This introduces the concept of full network nonlocality (FNN) \cite{Alej2022}. Non-bilocality and FNN have been verified via various experiments \cite{Carvacho2017,doi:10.1126/sciadv.1602743,PhysRevA.95.062315,Sun2019,Poderini2020,PhysRevLett.129.030502,PhysRevLett.130.190201,Wang2023}. All of these experiments have only considered networks consisting of all bipartite entangled two-photon states. While the assumption for replacing the multipartite quantum source in a network with a classical source is too strong, it intrigues a natural question of how to characterize the new nonlocality to rule out biseparable sources \cite{Svetlichny}.

\begin{figure}[!t]
	\centering
	\includegraphics[width=0.9\textwidth]{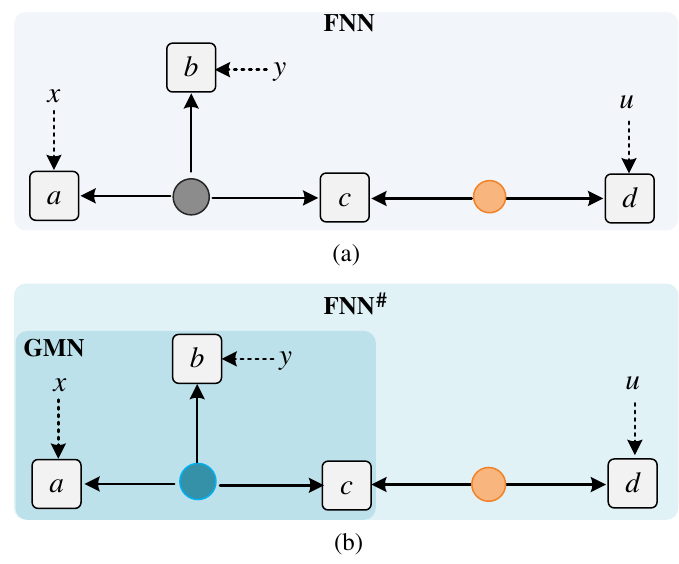}
	\caption{A causal network with four outcomes (gray box) of four observers, Alice, Bob, Charlie, and Debby, depends on (black arrows) the input states from one or two sources (circles) and the measurement settings ($x,y,u$).  Charlie performs a joint measurement on his systems yielding outcomes $c$. (a) Full network nonlocality (FNN), where FNN rules out the correlations from the hybrid network consisting of an LHV source (gray circle) and a general nonlocal source (purple circle). (b) Simultaneous strong FNN (FNN$^\#$) and genuine multipartite nonlocality (GMN), where FNN$^\#$ rules out the correlations from the hybrid model consisting of a biseparable source (blue circle) and a general nonlocal source (orange circle), and GMN rules out all biseparable correlations (deep blue square).} 
	\label{Figure1}
\end{figure}

In this paper, we propose a novel method to verify both nonlocalities in a single experiment simultaneously. We examine a quantum network comprising a bipartite entangled state and a tripartite genuinely entangled state, as depicted in Fig.~\ref{Figure1}. We demonstrate that local measurements for each observer reveal quantum correlations that cannot be simulated if the tripartite sources distribute biseparable systems \cite{Svetlichny}, while bipartite sources distribute even stronger-than-quantum systems. The new experiment makes two significant contributions. First, it verifies the GMN of generalized GHZ states, which cannot be verified using the standard Svetlichny inequality \cite{Svetlichny}, in a more accessible way with the assistance of entanglement swapping. Second, it introduces a new type of nonlocality in network scenarios. Since our model involves local biseparable and no-signaling sources, which is stronger than the mixture of bipartite classical and no-signaling sources \cite{Alej2022}, our findings reveal stronger nonlocality than what can be verified by the FNN framework. 

{\it Simultaneously Verifying GMN and strong FNN.--}Consider a local network model consisting of a biseparable source and a bipartite no-signaling source, as shown in Fig.~\ref{Figure1}. Here, Alice, Bob, and Charlie share one source while Charlie and Debby share one source. The outcome denoted by $a$ of Alice depends on the local measurement setting $x$ and its shares. Similarly, the outcome $b$ of Bob depends on the local measurement setting $y$ and its shares. The outcome $d$ of Debby depends on the local measurement setting $u$ and its shares. There is one measurement setting for Charlie with measurement outcome $c$. The goal here is to verify two kinds of nonlocalities simultaneously within one network Bell experiment.

One is to verify the joint correlations generated by Alice, Bob, and Charlie exhibit GMN to rule out those that can be simulated by the biseparable model \cite{Svetlichny}. There are also definitions of the GMN in the resource framework \cite{Gallego2012,Xavier2021}, whereas our discussion of GMN focuses on the partitions of correlations. Specifically, the tripartite joint correlations (e.g., $P_{bs}(a,b,c|x,y)$, without the assumption of measurement input for Charlie) generated when the shared source is biseparable allow the following decomposition \cite{Svetlichny}:
\begin{eqnarray}
P_{bs}(a,b,c|x,y)&=&p_1p(a|x)p(b,c|y)+p_2p(b|y)p(a,c|x)
\nonumber
\\
&&+p_3p(c)p(a,b|x,y),
\end{eqnarray}
where $p_i$ represents the probability distribution characterizing the classical mixture of biseparable correlations among Alice, Bob, and Charlie. Additionally, $p(a|x)$, $p(b|y)$, and $p(c)$ represent the probability distribution of the outcomes of Alice, Bob, and Charlie, respectively. $p(a,b|x,y)$, $p(b,c|y)$ and $p(a,c|x)$ are the joint probability distributions of the outcomes of Alice and Bob, Bob and Charlie, and Alice and Charlie, respectively. These joint probability distributions satisfy the single restriction of the no-signaling principle \cite{Popescu1992}, i.e.,  $\sum_s p(s,t|u,v)=p(t|v)$ and $\sum_t p(s,t|u,v)=p(s|u)$. 

The other is to verify the joint correlations of all parties are strong full network nonlocal (FNN$^\#$). The FNN correlations here are defined analogously to standard nonlocal ones \cite{Alej2022}, which do not admit the specific hybrid model shown in Fig.~\ref{Figure1}(a), where there is at least one classical source (left). With local measurements, all parties can generate joint correlations that allow the following decomposition as 
\begin{equation}   P(a,b,c,d|x,y,u)=\int d\lambda\mu(\lambda)p(a|x,\lambda)p(b|y,\lambda)p(c,d|u,\lambda),
\end{equation}
 where $\mu(\lambda)$ is the probability distribution characterizing the local measurable variable $\lambda$ shared by Alice and Bob and Charlie, $p(a|x,\lambda)$ is Alice's response function, $p(b|y,\lambda)$ is Bob's response function, and $p(c,d|u,\lambda)$ is a joint response function for Charlie and Debby, which satisfies the no-signaling principle, i.e., $\sum_c p(c,d|u,\lambda)=p(d|u)$ and $\sum_d p(c,d|u,\lambda)=p(c|\lambda)$. The FNN can be established by ruling out all the cases by interchanging sources. The strong FNN correlations (FNN$^\#$) are defined based on the model shown in Fig.~\ref{Figure1}(b), which is analogous to fully network nonlocal correlations \cite{Alej2022} ruling out correlations from networks consisting of at least one classical source and others being no-signaling sources \cite{Popescu1992}. To witness FNN$^\#$, we apply different network realizations, i.e., a network consisting of a tripartite biseparable source \cite{Svetlichny} (left) and one bipartite no-signaling source \cite{Popescu1992} (right). This allows the decomposition of the joint network local correlations as 
\begin{eqnarray}
P(a,b,c,d|x,y,u)=P_{bs}(a,b,c|x,y)P_{ns}(c,d|u),
\end{eqnarray}
with biseparable correlations $P_{bs}(a,b,c|x,y)$ and no-signalling correlations $P_{ns}(c,d|u)$. So far, there is no result in exploring the non-classicality of this kind of network.

In what follows, consider a Bell experiment on the network shown in Fig.~\ref{Figure1}(b). We show the non-classicality of quantum correlations generated by generalized entanglement-swapping \cite{Zukowski1993}, where we suppose that $a,b,d,x,y,u\in\{0,1\}$ and $c\in\{0,1,2\}$. All the joint correlations $P(a,b,c,d|x,y,u)$ generated from hybrid networks satisfy the following network Bell inequalities (see the proof in Supplemental Material):
\begin{equation}
\expe{S_4^{c}}/\expe{C^{c}}\leq 4, c=0,1,2,
    \label{lbilocal}
\end{equation}
where $S_4$ is a combination of a tripartite Svetlichny operator with a positive-operator valued measurement defined by 
\begin{eqnarray}
    S_4^c&=&A_0B_0C^cD_1+ A_0B_1C^cD_0+ A_1B_0C^cD_0
\nonumber\\
&&
-A_1B_1C^cD_1+A_1B_1C^cD_0+ A_1B_0C^cD_1
\nonumber\\
&&
+ A_0B_1C^cD_1 -A_0B_0C^cD_0,
\label{correlator}
\end{eqnarray}
and the mean value of the correlators $\expe{A_xB_yC^cD_z}$ is defined by $\expe{A_xB_yC^cD_z}=\sum_{a,b,d=0,1}(-1)^{a+b+d}P(a,b,c,d|x,y,z)$ when three leaf nodes perform Pauli measurement, and $\expe{C^c} = P(c)$ is the probability that Charlie obtains the output $c$. The inequalities (\ref{lbilocal}) mean that the three leaf nodes perform the Svetlichny test in the case where Charlie obtains the output $c$. Thus, we can examine the average result of the three Svetlichny tests in terms of the output probability obtained by Charlie:
\begin{equation}
    \expe{S_4^{0}} + \expe{S_4^{1}} + \expe{S_4^{2}}\leq 4, 
    \label{mean}
\end{equation}
The inequalities (\ref{lbilocal}) and (\ref{mean}) are further applied to witness both the GMN of partial tripartite correlations and FNN$^\#$ of four-partite joint correlations, i.e., any violation of them certifies both the GMN of tripartite source and the FNN$^\#$ of hybrid networks. 

In quantum scenarios, the network Bell inequalities (\ref{lbilocal}) and (\ref{mean}) can be violated by quantum correlations. Consider that the tripartite source emits a generalized GHZ state \cite{GHZ}: $\ket{GHZ(\theta)}=\cos\theta\ket{000}+\sin\theta\ket{111}$ and the bipartite source emits a generalized EPR state \cite{EPR}:  $\ket{EPR(\theta)}=\cos\theta\ket{00}+\sin\theta\ket{11}$ with $\theta\in (0,\frac{\pi}{2})$. Charlie performs the partial Bell state measurement (PBSM) under the basis $\{\frac{1}{\sqrt{2}}(\ket{01}+\ket{10}), \frac{1}{\sqrt{2}}(\ket{01}-\ket{10}), \ket{00}\bra{ 00}+\ket{11}\bra{11}\}$ labeled by $c = \{0,1,2\}$. When Charlie has an output of $c = 0$, Alice measures the observables $A_0=X$ and $A_1=Y$, Bob measures the observables  $B_0=-\frac{1}{\sqrt{2}}(X+Y)$ and $B_1=\frac{1}{\sqrt{2}}(X-Y)$, and Debby measures the observables $D_0=X$ and $D_1=-Y$, where $X$ and $Y$ are Pauli operators. The joint conditional distribution $P(a,b,c=0,d|x,y,u)$ in this scenario gives $\expe{S_4^{0}}/\expe{C^{0}} = 4\sqrt{2}$. When Charlie has an output of $c = 1$, Alice and Bob measure the same observables, while Debby measures the negative of the previous observables, i.e. $D_0 = -X$ and $D_1 = Y$. The joint conditional distribution $P(a,b,c=1,d|x,y,u)$ in this scenario gives $\expe{S_4^{1}}/\expe{C^{1}} = 4\sqrt{2}$. Both scenarios violate the inequality (\ref{lbilocal}), thus verifying the GMN of generalized GHZ states and FNN$^\#$ simultaneously. For the output of $c = 2$, the three leaf nodes become separable, in which case they all measure the identity operator. The joint conditional distribution $P(a,b,c=2,d|x,y,u)$ gives $\expe{S_4^{2}}/\expe{C^{2}} = 4$, which equals the classical bound. Examining the average of these three cases implies 
\begin{equation}
\expe{S_4^{0}} + \expe{S_4^{1}} + \expe{S_4^{2}} = 4 + 2(\sqrt{2}-1)\sin^2{2\theta}. 
    \label{meanvio}
\end{equation}
The quantum correlations generated from the hybrid network can always violate the inequality (\ref{mean}) regardless of the value of $\theta$. In contrast, the standard Svetlichny inequality can only certify the GMN for generalized GHZ states with $\theta > \pi/8$ (see Fig.~\ref{fig:result}). 

As all the observables share the same eigenbases and differ only in the sign of their eigenvalues, no feedback is required for this experiment test, i.e., two measurement inputs for each leaf node are sufficient to examine these network Bell inequalities. We extend to verify the general GMN of multipartite GHZ states and FNN$^\#$ of quantum networks in Supplemental Material. 

\begin{figure} 
    \centering
    \includegraphics[width = \textwidth]{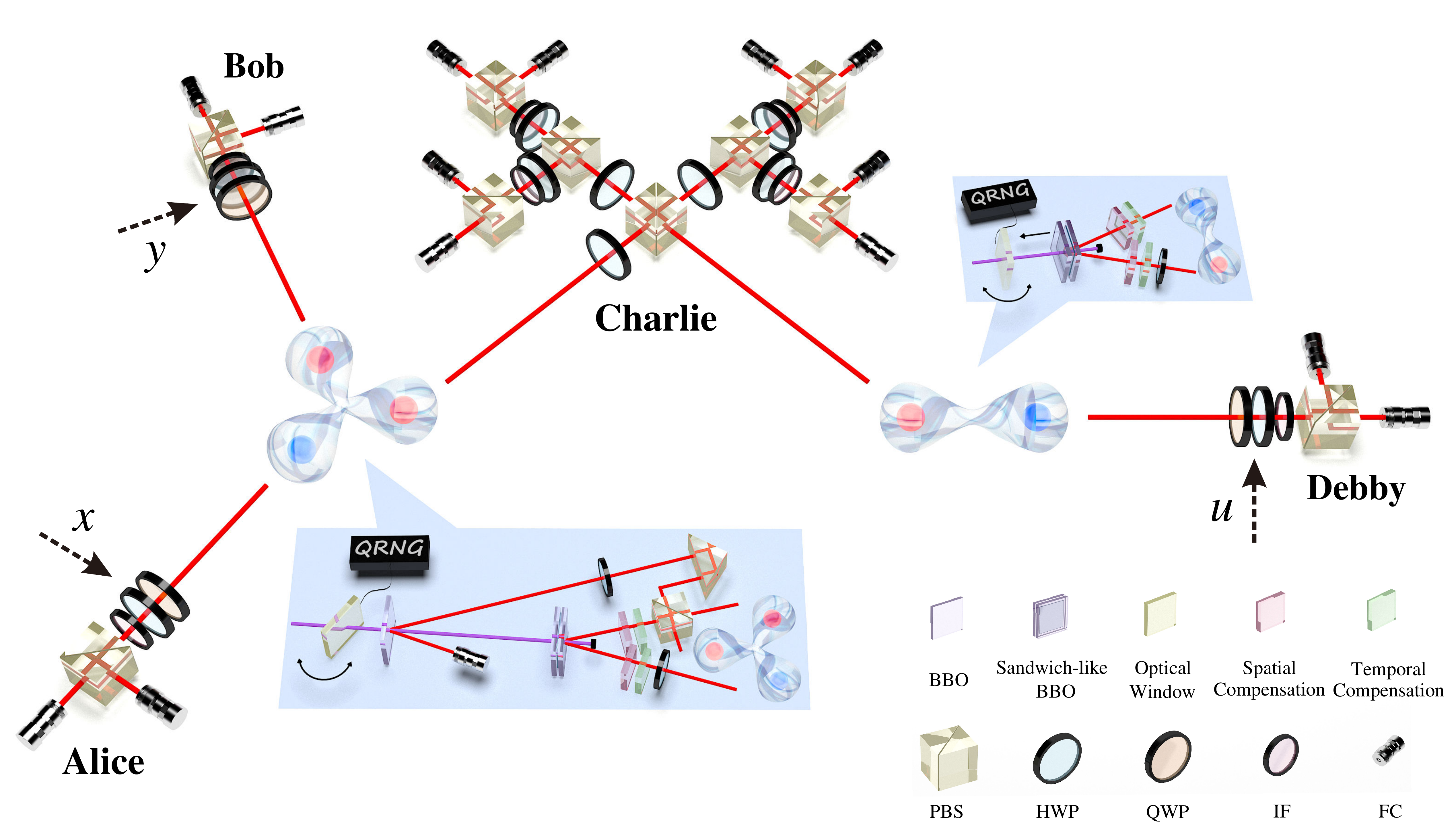}
    \caption{Sketch of the experimental setup. The network is composed of bipartite and tripartite sources. The bipartite source is a sandwich-like BBO-HWP-BBO structure, where the first BBO can be translated along the pump path, so that the photons collected by single-mode fibers from this crystal can be tuned, enabling the distribution of generalized EPR states. The tripartite source distributes generalized GHZ states, which are prepared by Hong-Ou-Mandel interference of a maximally entangled state with a single-qubit superposition state. The coherence between pump pulses is erased by randomly tilted optical windows. Charlie in the central node performs partial Bell state measurement using pseudo-number-resolving detectors. The other three nodes perform single-qubit measurements by setting the HWP and QWP to appropriate angles. BBO beta barium borate, PBS polarizing beam splitter, HWP half-wave plate, QWP quarter-wave plate, IF interference filter, FC fiber coupler, QRNG quantum random number generator. }
    \label{fig:setup}
\end{figure}

{\it Experiment.--}The quantum network is built on a photonic platform, as shown in Fig.~\ref{fig:setup}. We encode qubits using photonic polarization degrees of freedom, i.e., $\ket{0}$/$\ket{1}$ by the horizontally and vertically polarized states $\ket{H}$/$\ket{V}$. Photons are produced by the spontaneous parametric downconversion (SPDC) process in beta barium borate (BBO) crystals, which are pumped by ultraviolet pulses. These pulses, generated by a single laser, are split to pump two separate sources. To enhance the independence between the two sources, we insert a randomly tilting optical window controlled by a quantum random number generator into each pump path, which can apply a randomly varying phase to each pump beam, thus erasing the coherent information between the two beams. We measured the mutual information between any two leaf nodes, and the results were very close to 0 regardless of which measurement setting was chosen, aligning with the independent source condition (see Supplemental Material for more details). 

Our bipartite and tripartite photon sources can distribute generalized EPR states and generalized GHZ states, respectively, with $\theta$ tunable between 0 and $\pi/2$ (see Fig.~\ref{fig:setup} and Supplemental Material for more details). During the experiment, we set $\theta$ of each state to be as small as possible while keeping the success probability of the Bell states projection in the central node not too low. Through quantum state tomography, we find that the highest fidelity of the bipartite source compared to the generalized EPR state is attained at $\theta = 0.2734 \pm 0.0001$, which is $0.9809 \pm 0.0001$. We find that the highest fidelity of the tripartite source compared to the generalized GHZ state is given at $\theta = 0.2741 \pm 0.0014$, which is $0.9605 \pm 0.0011$. However, when measuring the standard Svetlichny inequality, we obtain a result of $2.7353 \pm 0.0172$, which does not exceed the classical bound 4, so the GMN of the tripartite source cannot be verified. 

In the hybrid network, each of the three leaf nodes receives one photon, and we employ waveplates and polarizing beam splitter (PBS) to conduct single-qubit Pauli measurements according to their inputs. The central node receives one photon from each source and performs partial PBSM, resulting in three outputs. The PBSM is realized with PBS and several pseudo-number-resolving detectors. The response of different groups of detectors distinguishes the three outputs (see Supplemental Material for more details). To characterize the setup, we perform the measurement tomography on our setup, and the result shows a fidelity of $0.9581 \pm 0.0005$ compared to the ideal PBSM. 

\begin{figure}
    \centering
    \includegraphics[width = \textwidth]{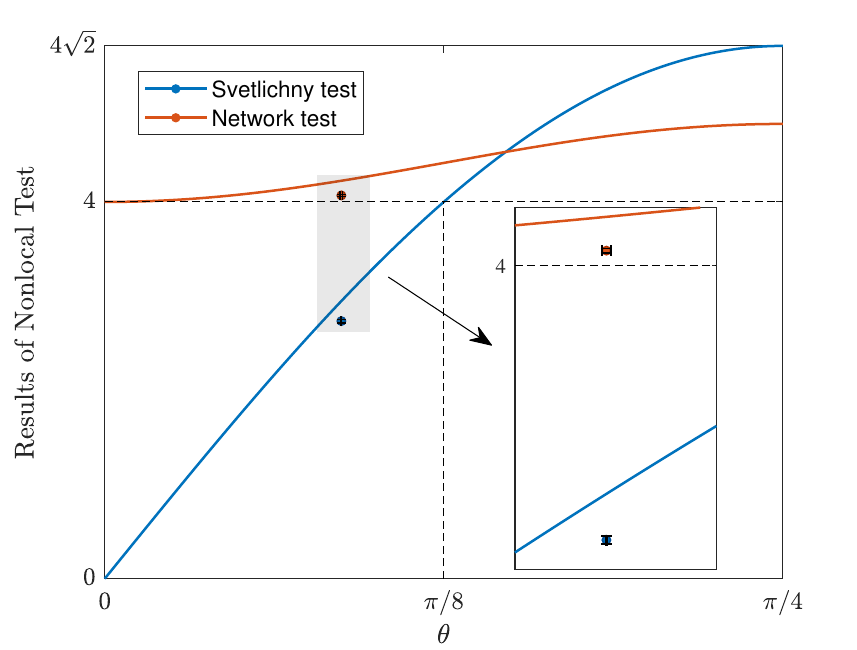}
    \caption{The experimental results of two nonlocal tests. The tripartite correlation generated by the generalized GHZ state cannot be certified as genuine multipartite nonlocality by violating the standard Svetlichny inequality when $\theta \le \pi/8$. In the hybrid network, the genuine multipartite nonlocality and strong full network nonlocality can be simultaneously verified by our network Bell inequality, regardless of $\theta$. The error bars are deduced from the photon statistical error of the raw data.}
    \label{fig:result}
\end{figure}

We experimentally performed 8 sets of measurement settings, each collecting approximately 5300 six-photon coincidence events over 60,000 seconds. By normalizing the raw data we can obtain the distribution $P(a,b,c,d|x,y,u)$ (see Supplemental Material for more details). Therefore, we find $\expe{S_4^{0}}/\expe{C^{0}} = 4.4056 \pm 0.0868$, $\expe{S_4^{1}}/\expe{C^{1}} = 4.4443 \pm 0.0865$ and $\expe{S_4^{2}}/\expe{C^{2}} = 4$. By calculating the marginal probability we find $\expe{C^0} = 0.0815 \pm 0.0010$, $\expe{C^1} = 0.0815 \pm 0.0011$ and $\expe{C^2} = 0.0870 \pm 0.0015$. The average result of three inequalities consequently yielded $\expe{S_4^{0}} + \expe{S_4^{1}} + \expe{S_4^{2}} = 4.0693 \pm 0.0101$, surpassing the classical bound by more than 6 standard deviations (see Fig.~\ref{fig:result}). Thus, our results simultaneously witness GMN of the tripartite source and FNN$^\#$ of the quantum network.

{\it Conclusion.--}The present scheme is robust against noise. Consider a noisy network consisting of two Werner states \cite{Werner}: $\rho_v=v_1\ket{GHZ(\theta)}\bra{GHZ(\theta)}+\frac{1-v_1}{8}\mathbbm{1}_8$ and $\varrho_v=v_2\ket{EPR(\theta)}\bra{EPR(\theta)}+\frac{1-v_2}{4}\mathbbm{1}_4$, where $v_i\in (0,1]$ denotes the noise parameters of two entangled states and $\mathbbm{1}_k$ is the identity matrix with rank $k$. The linear correlator (\ref{correlator}) implies a total noise visibility of $v_1v_2\geq \frac{1}{\sqrt{2}}$, which further allows a single constant noise visibility of $v_1\geq \frac{1}{\sqrt{2}}$ for witnessing all generalized GHZ pure states. This implies that any nontrivial quantum correlations activated from entanglement swapping show the non-classicality of both quantum entanglement, going beyond the standard Bell test by using the MABK inequalities, Svetlichny inequality \cite{Mermin,Ardehali,Belinskii,Svetlichny,Ajoy2010} or the Hardy inequality \cite{Chen,Yu}. Similar results hold for witnessing all generalized EPR states beyond the CHSH test \cite{CHSH}. This means that the present network method shows a new possibility for witnessing noisy entanglement beyond Bell experiments with single entanglement. The present scheme generally does not require the consistency of two entangled sources. The extended setting up, i.e., $\cos\theta_1\ket{000}+\sin\theta_1\ket{111}$ and $\cos\theta_2\ket{00}+\sin\theta_2\ket{11}$ with $\theta_1\not=\theta_2$, requires a generalized Bell-basis measurement performed by Charlie. Our results provide new insight into quantum networks and intrigue further studies to explore the non-classicality of quantum networks with different resources.  

Experimentally, we verified FNN$^\#$, which can rule out network correlations generated by at least one biseparable source, in a high-fidelity photonic hybrid network consisting of tripartite and bipartite sources, employing entanglement swapping. Simultaneously, the results verified the GMN exhibited by the generalized GHZ states, even if it cannot be directly verified via the standard Svetlichny inequality \cite{Svetlichny}. Thus, two kinds of complex nonlocalities are verified in a single experiment. This approach can also be viewed as a network-assisted device-independent certification protocol for genuine multipartite entanglement \cite{PhysRevLett.129.190503,supic2023}. Finally, we note that our results are subject to common loopholes in nonlocal experiments: locality loophole, source independence loophole, detection loophole, and freedom-of-choice loophole. Further experiments conducted on larger spatial scales \cite{Sun2019, PhysRevLett.130.190201} and utilizing herald setups \cite{PhysRevLett.132.130604, PhysRevLett.132.130603} offer possible routes for closing these loopholes. 

{\it Acknowledgments.--} This work was supported by the National Natural Science Foundation of China (Nos.62172341, 62075208, 11821404, 11734015), the Fundamental Research Funds for the Central Universities (No. YD2030002015), Sichuan Natural Science Foundation (Nos.2023NSFSC0447,2024NSFSC1365,2024NSFSC1375), Innovation Program for Quantum Science and Technology (No. 2021ZD0301604), Interdisciplinary Research of Southwest Jiaotong University China (No.2682022KJ004). This work was partially carried out at the USTC Center for Micro and Nanoscale Research and Fabrication. 


%

\end{document}